\documentclass[sigconf]{acmart}
\AtBeginDocument{%
  }

\copyrightyear{2025} 
\acmYear{2025} 
\setcopyright{acmlicensed}\acmConference[SIGIR '25]{Proceedings of the 48th International ACM SIGIR Conference on Research and Development in Information Retrieval}{July 13--18, 2025}{Padua, Italy.}
\acmBooktitle{Proceedings of the 48th International ACM SIGIR Conference on Research and Development in Information Retrieval (SIGIR '25), July 13--18, 2025, Padua, Italy}
\acmDOI{10.1145/3726302.3730046}
\acmISBN{979-8-4007-1592-1/2025/07}

\usepackage{multirow}
\usepackage{graphicx} 
\usepackage{colortbl}
\usepackage{balance}
\usepackage{enumitem}
\usepackage{subfigure}
\newcommand\ie{{\it i.e.}}
\newcommand\eg{{\it e.g.}}
\begin{document}

\title{NR4DER: Neural Re-ranking for Diversified Exercise Recommendation}

\author{Xinghe Cheng}
\orcid{0000-0001-9432-5794}
\affiliation{
  \institution{Jinan University}
  \city{Guangzhou}
  \country{China}
}
\email{jnuchengxh@hotmail.com}

\author{Xufang Zhou}
\orcid{0009-0008-7222-1466}
\affiliation{
  \institution{Jinan University}
  \city{Guangzhou}
  \country{China}}
  \email{etally@stu.jnu.edu.cn}
  
\author{Liangda Fang}
\orcid{0000-0002-6435-6570}
\authornote{Corresponding Author.}
\affiliation{
  \institution{Jinan University\\ Pazhou Lab}
  \city{Guangzhou}
  \country{China}}
  \email{fangld@jnu.edu.cn}
  
\author{Chaobo He}
\orcid{0000-0002-6651-1175}
\affiliation{
 \institution{South China Normal University}
  \city{Guangzhou}
  \country{China}}
  \email{hechaobo@foxmail.com}

\author{Yuyu Zhou}
\orcid{0009-0000-0611-9449}
\affiliation{
  \institution{Jinan University}
  \city{Guangzhou}
  \country{China}}
  \email{zyy@jnu.edu.cn}
  
\author{Weiqi Luo}
\orcid{0000-0001-5605-7397}
\affiliation{
  \institution{Jinan University}
  \city{Guangzhou}
  \country{China}}
  \email{lwq@jnu.edu.cn}

\author{Zhiguo Gong}
\orcid{0000-0002-4588-890X}
\affiliation{
  \institution{University of Macau}
  \city{Macau}
  \country{China}}
  \email{fstzgg@um.edu.mo}

\author{Quanlong Guan}
\orcid{0000-0001-6911-3853}
\authornote{Corresponding Author.}
\affiliation{
  \institution{Jinan University}
  \city{Guangzhou}
  \country{China}}
  \email{gql@jnu.edu.cn}

\renewcommand{\shortauthors}{Xinghe Cheng et al.}

\begin{abstract}
	\looseness=-1
    With the widespread adoption of online education platforms, an increasing number of students are gaining new knowledge through Massive Open Online Courses (MOOCs).
    Exercise recommendation have made strides toward improving student learning outcomes.
    However, existing methods not only struggle with high dropout rates but also fail to match the diverse learning pace of students.
    They frequently face difficulties in adjusting to inactive students' learning patterns and in accommodating individualized learning paces, resulting in limited accuracy and diversity in recommendations.
    To tackle these challenges, we propose Neural Re-ranking for Diversified Exercise Recommendation (in short, NR4DER).
    NR4DER first leverages the mLSTM model to improve the effectiveness of the exercise filter module.
    It then employs a sequence enhancement method to enhance the representation of inactive students, accurately matches students with exercises of appropriate difficulty.
    Finally, it utilizes neural re-ranking to generate diverse recommendation lists based on individual students' learning histories.
    Extensive experimental results indicate that NR4DER significantly outperforms existing methods across multiple real-world datasets and effectively caters to the diverse learning pace of students.
\end{abstract}




\maketitle

\section{Introduction}
\looseness=-1
With the rapid advancement of internet technologies, an increasing number of students are turning to online education platforms, such as Massive Online Open Courses (MOOCs) \cite{FenTL2019}, with the aim of acquiring knowledge, strengthening their skills, and exploring new subjects.
Exercise recommendation is essential in assisting students in acquiring knowledge by providing appropriate exercises rather than leaving them to conduct self-directed searches \cite{UrdMO2021}.
Despite the significant achievements of existing exercise recommendations, it still struggles to address long-tailed student distribution problem \cite{FanLW2021,JanLC2020} and capture the diverse learning pace of students \cite{WonL2020, BurP2013}.
    \begin{figure}[t]
	\centering
	\includegraphics[width= 1\linewidth]{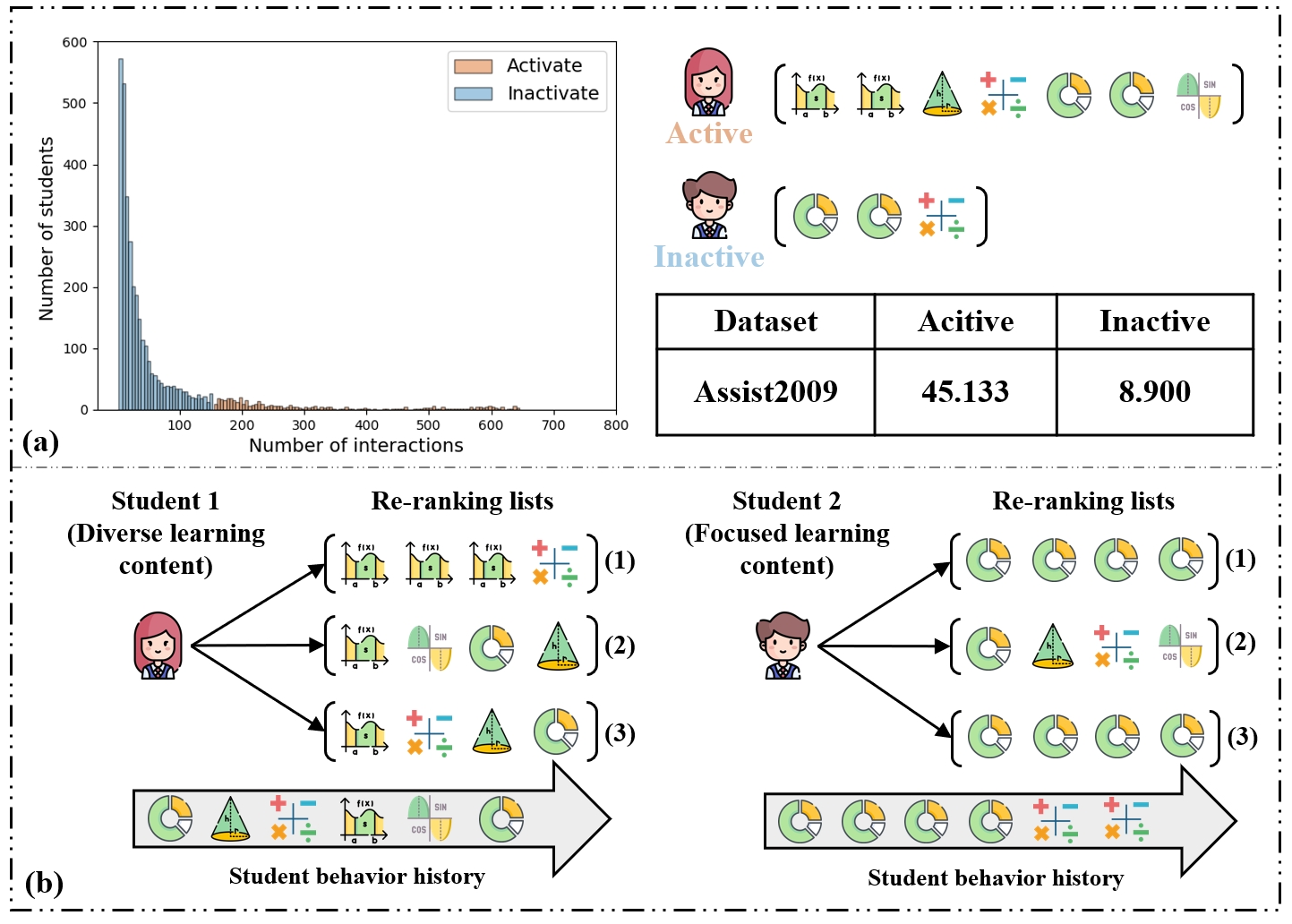}
	\caption{(a) Distribution of interaction frequencies and diversity metric in datasets, we categorizing students into active students (top 5\% interactions, orange) and inactive students (bottom 95\% interactions, blue). (b) An example of different re-ranking strategies.}
        \Description{background}
	\label{fig:background}
    \end{figure}

\looseness=-1
Currently, online education platforms face the challenge of high dropout rates \cite{MolOB2022, BorCP2022}, resulting in the long-tailed student distribution problem \cite{LiuZ2020,YunKY2022}.
This problem, prevalent in online education, arises when students with few interactions (i.e., inactive students) significantly outnumber those with many interactions (i.e., active students). \cite{LiuFW2021, YinLW2020}.
As shown in Figure \ref{fig:background}(a), the number of active students with sufficient and diverse historical exercise records is relatively small, while the majority of inactive students have fewer and more monotonous exercise records.
We observe that active students achieved a significantly higher diversity score of 45.133 compared to merely 8.900 for inactive students.
Hence, training on long-tailed data usually results in a significant bias towards classes with sufficient records (activate students) and performs poorly on rare classes (inactivate students), giving rise to the long-tailed student distribution problem.

\looseness=-1
On the other hand, a few research has been undertaken in the realm of exercise recommendation to capture the diverse learning pace of students.
Exercise recommendation is categorized into three types: Knowledge Tracing (KT)-based \cite{HuoWN2020}, Reinforcement Learning (RL)-based \cite{HuaLZ2019, LiuHL2023}, and Multi-Stage (MS)-based approaches \cite{WuLT2020, GuaXC2023, RenLS2023}.
KT-based and RL-based exercise recommendation primarily focus on the knowledge concepts that students do not master \cite{HuoWN2020, LiuHL2023}.
Although these approaches have achieved excellent performance, they are often limited by local greedy optimality, thus recommend homogeneous exercises. 
MS-based exercise recommendations contains three stages: recall, ranking, and re-ranking.
The re-ranking stage directly influences the student learning process and effectiveness and plays a vital role in recommendations.
They aim to maximize the diversity of exercises related to different knowledge concepts in the recommendation list \cite{WuLT2020, RenLS2023}.
However, most MS-based methods often lead to similar re-ranked results and neglect students' learning pace for diversity \cite{LiuHL2023}.
As shown in Figure \ref{fig:background}(b), some students tend to master a specific set of knowledge concepts through practicing, while others prefer to cover a broader range.
This poses a challenge for exercise recommendation methods, making it difficult to effectively recognize and match the diverse learning pace of students \cite{SonWW2021}.

\looseness=-1
To overcome the aforementioned two problems, we develop a \textbf{N}eural \textbf{R}e-ranking approach for \textbf{D}iversified \textbf{E}xercise \textbf{R}ecommend-ation (in short, NR4DER).
The main contributions are the following.
\begin{enumerate}
	\item We first construct an exercise filter module to recommend appropriate difficulty exercises for different students. 
	Additionally, to tackle the long-tailed student distribution problem, we propose a student representation enhancer that leverages the rich historical learning record of active students to improve overall performance.
	
	\item We adopt the neural re-ranking proposed by \citet{LiuXQ2023} in e-commerce recommendation to enhance diversity in educational recommendation scenarios. This method generates diverse re-ranked lists tailored to each student's learning pace.
	
	\item We conduct extensive experiments on three real-world datasets to validate the effectiveness of NR4DER in terms of both accuracy and diversity.
\end{enumerate}

\section{Related Work}
\subsection{Exercise Recommendation}
\looseness=-1
Exercise recommendation (ER) systems analyze the learning states of students to recommend suitable exercises that improve comprehension of course content.
Existing approaches can be classified into three main categories: KT-based, MS-based and RL-based approaches.
In KT-based methods, some studies utilize KT to assess students' mastery and provide tailored recommendations \cite{LiuHL2023}.
LSTMCQP \cite{HuoWN2020} introduced two contextualized representations to enhance LSTM-based knowledge tracing.
In RL-based methods, DRER \cite{HuaLZ2019} implements a novel deep reinforcement learning framework to optimize multi-objective recommendations in online education.
MMER \cite{LiuHL2023} incorporates a multi-agent module that treats knowledge concepts as both competing and cooperating, thereby enhancing the robustness of recommendation for new students through meta-training.
However, many existing models only focus on students' mastery of knowledge concepts and neglect some metrics such as diversity, which can lead to recommendations that fail to meet students' needs. 
In MS-based methods, KCPER \cite{WuLT2020} combines LSTM with DKT to predict students' knowledge states and recommend tailored exercises while using a simulated annealing algorithm to enhance diversity.
Due to the interpretability of knowledge graphs \cite{WanKL2024, WanWG2024, WanCW2024}, both KG4Ex \cite{GuaXC2023} and KG4EER \cite{GuaCX2025} employ interpretability via constructing a knowledge graph that illustrates the relationships among knowledge concepts, students and exercises.
Nevertheless, existing exercise recommendation methods do not fix the long-tailed student distribution problem, resulting in a significant bias towards classes with sufficient records (active students) and poor performance on classes with sparse records (inactive students).

\looseness=-1
To address the long-tailed student distribution problem, we introduce the student representation enhancer to enrich inactive student representations.
Additionally, to tackle the insufficiency of diversity in exercise recommendations, we incorporate neural re-ranking, which more accurately captures students' learning pace while balancing knowledge relevance and diversity.

\subsection{Neural Re-ranking}
\looseness=-1
In recent years, neural re-ranking has gained significant prominence across recommendation scenarios.
These approaches utilize neural networks to perform fine-grained ranking of candidate items, thereby enhancing the quality of recommendations.

\looseness=-1
Neural re-ranking can be categorized into two types: single-objective and multi-objective.
Single-objective re-ranking can be learned through either observed signals or counterfactual signals.
Observed signal-based methods leverage user behavior data (\eg, clicks and favorites) as supervision signals to train neural networks for re-ranking \cite{ZhuOW2018, CarG1998}.
In contrast, counterfactual signal-based learning addresses the sparsity of observational data. 
SEG \cite{WanFL2019} uses a generator-evaluator framework and reinforcement learning to generate rankings that align with user preferences, while CDIA \cite{PanXA2020} employs an actor-critic framework to optimize the model.
However, these single-objective approaches primarily focus on accuracy and often overlook the need for diversity in recommendations. 
Multi-objective re-ranking, while ensuring accuracy, also emphasizes diversity. $\mathrm{M}^2\mathrm{DIV}$ \cite{FenXL2018} combines reinforcement learning to achieve diversity-aware re-ranking, and RAPID \cite{LiuXQ2023} analyzes users' historical behavior to learn their preferences for different topics, balancing relevance and diversity.
However, these methods cannot be directly applied to the exercise recommendation.

\looseness=-1
Hence, we propose NR4DER, which integrates neural re-ranking with an exercise recommendation algorithm that takes into account knowledge states of students.
This approach further explores the relationships between knowledge concepts to generate diverse exercise recommendation lists for students.

    \begin{figure*}
        \includegraphics[width=0.9\textwidth]{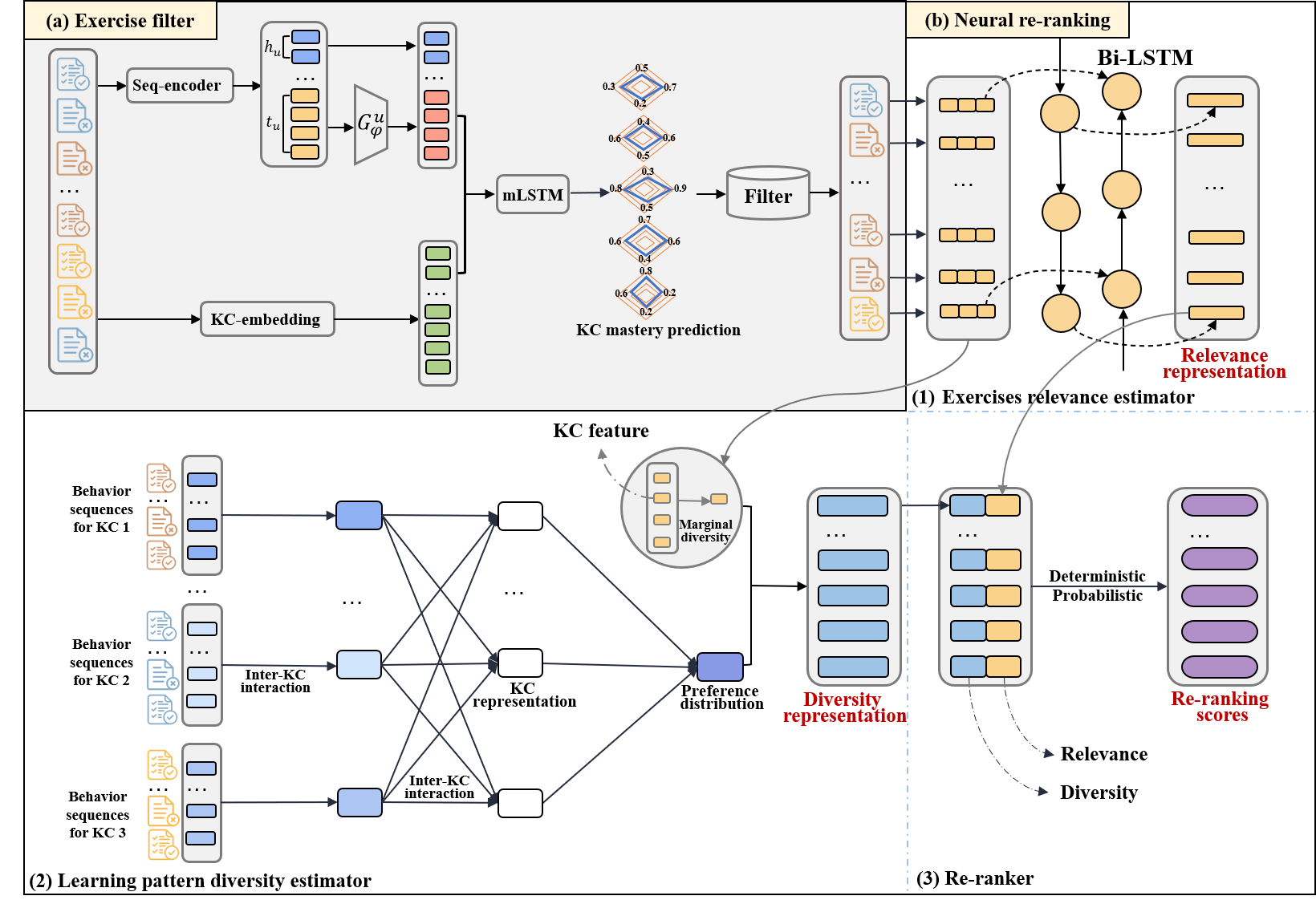}
        \caption{NR4DER Framework. The exercise filter module (a) takes students' historical exercise sequence as input and generates a list of candidate exercises with suitable difficulty for each student.
        The neural re-ranking module (b) first processes the embeddings of exercises and knowledge concepts to compute relevance and diversity matrices, then computes re-ranking scores and finally recommends exercises based on these scores.
        This module consists of three key components: exercise relevance estimator, learning pattern diversity estimator, and re-ranker.}
        \Description{work flow chart}
        \label{fig:work flow chart}   
	\end{figure*}

\section{Problem Definition and Preliminaries}
In this section, we provide the definition of exercise recommendation and introduce some preliminaries about  mLSTM and exercise difficulty estimation.

\subsection{Problem Definition}
\looseness=-1
In an online education platform, two sets $S$ of students and $E$ of exercises are given.
We use $|S|$ and $|E|$ for the number of students and exercises, respectively.
The exercising process of student $s \in S$ is defined as a sequence $P_s = \{ (e_1, a_1), (e_2, a_2), \cdots , (e_t, a_t) \}$ where $e_i \in E$ represents the exercise that student $s$ practices at her time step $i$, and $a_i$ denotes the corresponding performance for $1 \leq i \leq t$.
More formally, if she answers exercise $e_i$ right, $a_i = 1$, otherwise $a_i = 0$.
For exercise $e \in E$, it is defined as a set of knowledge concepts $e = \{k_1, k_2, \cdots, k_m\}$.
A knowledge concept $k \in e$ means that exercise $e$ involves $k$.
Given an exercising process $\{ (e_1, a_1), (e_2, a_2), \cdots, (e_t, a_t ) \}$ of student $s$, exercise recommendation aims to output a suitable exercise $e_{t + 1}$ for student $s$.

\subsection{MLSTM and Exercise Difficulty Estimation}

\subsubsection{mLSTM}
\looseness=-1
Numerous methods have demonstrated that mLSTM \cite{BecPS2024,FanTZ2024} performs better in sequential tasks compared to RNN-based methods \cite{Hoc1997} and Transformer-based methods \cite{Vas2017}.
mLSTM utilizes the matrix memory to enhance storage capacity and handle long-term dependencies better than LSTM.
Meanwhile, mLSTM integrates the covariance update rule into the LSTM framework, where the forget gate corresponds to decay rate and the input gate to the learning rate, while the output gate scales the retrieved vector.
The covariance update rule for storing a key-value pair is defined as:
\begin{equation} \label{equation:mlstm}
        \begin{aligned}
            \mathbf{C}_t = \mathbf{f}_t \mathbf{C}_{t - 1} + \mathbf{i}_t \mathbf{v}_t \mathbf{k}_{t}^{\top}, \\
        \end{aligned}
    \end{equation}
where $\mathbf{f}_t$, $\mathbf{i}_t$, and $\mathbf{C}_t$ are forget gate, input gate, and cell state, respectively.
Furthermore, mLSTM can store a pair of vectors, the key $\mathbf{k}_t \in \mathbb{R}^d$ and the value $\mathbf{v}_t \in \mathbb{R}^d$.
Hence, mLSTM achieves parallelism by abandoning memory mixing, hence accelerating computation on GPUs.

\subsubsection{Exercise Difficulty Estimation}
\looseness=-1
The probability of a student solving an exercise correctly is commonly used to indicate the difficulty level of the exercise for the student \cite{WuLT2020}.
The former notion depends on the mastery of the knowledge concept involved in the exercise.
Since an exercise involves multiple knowledge concepts, we can use the following equation to obtain the difficulty of the exercise:
    \begin{equation} \label{equation:exercise difficulty}
        \begin{aligned}
             D_{e}=1-\prod_{k\in e}{\mathbb{P} (k)}
        \end{aligned}
    \end{equation}
where $D_{e}$ represents the difficulty of exercise $e$.
$\mathbb{P} (k)$ represents the level at which the student has mastered all knowledge concepts (\ie, $\mathbb{P} (k) = [ \mathbb{P} (k_1), \mathbb{P} (k_2),\cdots,\mathbb{P} (k_m)]$).
    
\section{Methodology}
In this section, we introduce our approach NR4DER shown in Figure \ref{fig:work flow chart}.
Our approach contains two key modules: the exercise filter module (Section 4.1) and the neural re-ranking module (Section 4.2).
The exercise filter module aims to filter exercise candidate subsets according to difficulty.
The neural re-ranking module aims to re-rank exercise candidate subsets to match the diverse learning pace of students.
Notations are described in Table \ref{tab:notations}.

\begin{table}[!t]
\centering
\caption{Notations}
\begin{tabular}{cl}
\hline
\textbf{Notation} & \textbf{Description} \\ \hline
$S$ & The set of students set \\
$E$ & The set of exercises set \\
$K$ & The set of knowledge concepts \\
$N$ & The number of students \\ 
$L$ & The number of exercises \\
$M$ & The number of knowledge concepts\\ 
$C$ & The candidate subset of exercise \\ 
$D$ & The difficulty of exercise \\ 
$\text{mLSTM}$ & mLSTM network \\ 
$\text{Attention}$ & Self-attention network \\ \hline
\label{tab:notations}
\end{tabular}
\end{table}

\subsection{Exercise Filter}
\looseness=-1
Figure \ref{fig:work flow chart}(a) shows the exercise filter module, which consists of student representation enhancer, knowledge concept mastery predictor and exercise filter.
To tackle the long-tailed student distribution problem, the student representation enhancer leverages the rich historical learning record of active students to improve overall performance.
The knowledge concept mastery predictor uses mLSTM model \cite{BecPS2024} to predict students' mastery of knowledge concept.
Finally, to filter exercise candidate subsets with suitable difficulty, we adopt an exercise filter to generate an exercise candidate subset.

\subsubsection{Student Representation Enhancer}
\looseness=-1
As learning is a continuous and dynamic process, the knowledge state of students evolves progressively.
The most recent interactions provides a more accurate reflection of their current learning progress, excluding outdated or irrelevant information and thereby enhancing the alignment of recommendations with their needs.
Hence, for an active student, its embedding representation obtained from entire exercise record is similar to that from the most recent interactions.
Inspired by this assumption, we adopt the approach proposed in \cite{KimHY2023} so as to acquire the informative representation of inactive students by making use of the most recent interactions of active students.
We first truncate the complete exercise sequence $P_s$ of active students into a subsequence $\bar{P}_s$ containing the most recent $T$ interactions:
    \begin{equation} \label{equation:subsequence}
        \begin{aligned}
             \bar{P}_s = [(e_{|s| - T + 1}, a_{|s|- T + 1}), \cdots, (e_{|s|}, a_{|s|})].
        \end{aligned}
    \end{equation}
Due to its short length, $\bar{P}_s$ can be considered as the exercise sequence of an inactive student and used to simulate scenarios where a student has limited interactions.

\looseness=-1
Subsequently, the model is trained to generate the representation $\mathbf{h}_s$ of the active student based on $\bar{P}_s$, and to transfer the acquired knowledge to inactive students.
We utilize an exercise sequence encoder $f(\cdot)$ to generate the representation of an active student based on a truncated exercise sequence $\mathbf{r}_s$:

    \begin{equation} \label{equation:sequence encoder}
        \begin{aligned}
             \mathbf{r}_s = f(\bar{P}_s) \in \mathbb{R}^d , 
        \end{aligned}
    \end{equation}
where $\mathbf{r}_s$ captures the learning pattern of student $s$.

\looseness=-1
Then, we generate a student representation that approximates the complete representation of active student $\mathbf{h}_s$ from $\mathbf{r}_s$ by minimizing the following loss:

    \begin{equation} \label{equation:enhancement loss function}
        \begin{aligned}
             \mathcal{L}_s = w_s |\mathbf{h}_s - G_{\phi}^{s}(\mathbf{r}_s)|^2, 
        \end{aligned}
    \end{equation}
where $G_{\phi}^{s}: \mathbb{R}^d \rightarrow \mathbb{R}^d$ denotes a student representation enhancer that generates the complete representation from the input $\mathbf{r}_s$ and $w_s$ denotes the loss coefficient for active student $s$ defined as

    \begin{equation} \label{equation:curriculum learning}
        \begin{aligned}
             w_s=\text{sin}( \frac{\pi}{2}\cdot (\frac{\text{epo}}{\text{epo}_{\text{max}}} + \frac{| p_s |-L_{\text{min}}}{L_{\text{max}}-L_{\text{min}}}) ) , 
        \end{aligned}
    \end{equation}
where $\text{epo}$ is the current epoch, $\text{epo}_{\text{max}}$ is the maximum number of epochs and $L_{\text{min}}$ and $L_{\text{max}}$ denotes the minimum and maximum of $|P_s|$ in the training data, respectively.
Active students with longer sequence lengths dominate the learning process during the early stages of training.
As training proceeds, however, it gradually shifts to learning more from active students with shorter sequence lengths \cite{LosH2016}.
Through minimizing Equation \eqref{equation:enhancement loss function}, $G_{\phi}^{s}$ acquires sufficient knowledge to reconstruct a complete student representation from a short exercise sequence, which can then be leveraged to enhance the representation of an inactive student $\mathbf{h}_{s}$ as follows:

    \begin{equation} \label{equation:enhance representation}
        \begin{aligned}
             \mathbf{h}_{s}^{+} = \begin{cases}
	G_{\phi}^{s}\left( \mathbf{r}_{s} \right) +\beta \mathbf{h}_{s}, & \hfill s \in S^I,\\
	\mathbf{h}_{s}, & \hfill \text{otherwise},\\
        \end{cases}
        \end{aligned}
    \end{equation}
where $\mathbf{h}_{s}^{+}$ is the enhanced representation of student $s$.
We remark that only the representations $\mathbf{h}_{s}$ of inactive students $s \in S^I$ are enhanced but those of active students $s \notin S^I$ are not.
$\mathbf{r}_{s} = f \left(P_s \right) \in \mathbb{R}^d$ denotes the representation of an inactive student $s$ obtained from her complete exercise sequence $P_{s}$. $\beta \in [0, 1]$ is a hyperparameter that controls the portion of the original representation $\mathbf{h}_{s}$.
The representation $\mathbf{h}_{s}$ of inactive students can be augmented as $\mathbf{h}_{s}^{+}$ by incorporating the representation based on the knowledge acquired from active students $G_{\phi}^{s}(\mathbf{r}_{s})$.

\subsubsection{Knowledge Concept Mastery Predictor}
\looseness=-1
The probability of correctly solving exercise reflects the student's mastery of the corresponding knowledge concepts.
We utilize mLSTM model \cite{BecPS2024} to predict student's knowledge concept mastery, named KCMP.

    \begin{equation} \label{equation:MDKT}
        \begin{aligned}
             y( k ) =\mathrm{mLSTM}( \mathbf{h}_{s}^{+} ),
        \end{aligned}
    \end{equation}
where $\mathbf{h}_{s}^{+}$ is the enhanced representation of student $s$ and $y(k)$ represents the mastery of the knowledge concept (\ie, $\mathbb{P} (k)$).

\looseness=-1
We denote the one-hot encoding of the knowledge concept answered at time $t+1$ by  $\phi(k^{t + 1})$ and the corresponding result by $a^{t+1}$.
The output of KCMP at time $t$ is denoted by $y^t$.
The optimization cost function of KCMP can be expressed as:

    \begin{equation} \label{equation:MKT loss}
        \begin{aligned}
             \mathcal{L}_K = \frac{1}{\sum_{i = 1}^N{T_i}} \sum_{i = 1}^N{\sum_{t = 0}^{T_i}{l((y_{i}^{t}) \cdot \phi(k^{t + 1}), a^{t + 1})}},
        \end{aligned}
    \end{equation}   
where $N$ is the number of students and $T_i$ indicates the time period for the $i$-th student to answer previous exercises.

\subsubsection{Model Training}
\looseness=-1
KCMP is trained based on the following loss containing parameters for the embedding generators and the exercise sequence encoder as follows:

    \begin{equation} \label{equation:final loss}
        \begin{aligned}
         \mathcal{L} _{\text{final}}=\lambda _s\sum_{s\in S^A}{\mathcal{L} _s}+\mathcal{L} _K,    
        \end{aligned}
    \end{equation}
where $\lambda_s$ is a hyperparameter for the loss $\mathcal{L} _s$ and $S^A$ is the set of active students.

\looseness=-1
After KCMP completes the training, we can predict the degree of a student mastering different knowledge concepts and compute the difficulty of each exercise for different students based on Equation \eqref{equation:exercise difficulty}.

\subsubsection{Exercise Filter}
\looseness=-1
The distance $\text{Dis}$ measures the distance between the prediction difficulty of exercise.
The parameter $\delta$ is the predefined difficulty threshold that used to filter exercises based on covered knowledge concepts.
Here, we get the difficulty of exercise by $D_{e}$ by Equation\eqref{equation:exercise difficulty}.
$\varPhi _{e}$ represents the weight of the exercise $e_i$, defined as:

    \begin{equation} \label{equation:difficulty distance}
        \begin{aligned}
             \varPhi _{e}=| \text{Dis}( \delta ,D_{e} ) |.
        \end{aligned}
    \end{equation}

\looseness=-1
Once the weight of each exercise $\varPhi _{e}$ is obtained, this module sorts the exercises in an ascending order of weights and selects $L$ exercises with the smallest weights as the an exercise candidate subset, denoted as $C$.

\subsection{Neural Re-ranking}
\looseness=-1
For a student $s \in S$ with a candidate subset $C_s$ of $L$ exercises, the neural re-ranking module seeks to learn a multivariate function over $C_s$, taking into account both exercise relevance and learning pattern diversity.
The exercises with the top-k highest re-ranking scores form the re-ranked list $R_s$, which is subsequently provided to the student with $K \leq L$.

\looseness=-1
Motivated by the approach of RAPID \cite{LiuXQ2023}, we adopt neural re-ranking to provide diverse re-ranked lists catering to the learning pace of each student in educational scenarios.
Figure \ref{fig:work flow chart}(b) shows the neural re-ranking module, which is consist of exercise relevance estimator, learning pattern diversity estimator and re-ranker.
The exercise relevance estimator uses the student, exercise, and knowledge concept features of the candidate subsets as input and then models the listwise context of the subsets.
By extracting the distribution of students' individual learning pace across various knowledge concepts from exercise records, the learning pattern diversity estimator integrates this distribution with knowledge concept coverage information to compute the learning pattern diversity gain of candidate exercises.
Finally, to predict the re-ranking scores in the re-ranker module, we adopt deterministic and probabilistic approaches to aggregate the estimated exercise relevance and learning pattern diversity.

\subsubsection{Exercise Relevance}
Let $C_l$ denote the $l$-th exercise in the candidate subsets $C$ where $1 \leq l \leq L$.
For exercise $C_l$, we represent it by integrating three components: the student feature $\mathbf{x}_s$, the exercise feature $\mathbf{x}_{C_l}$, and the knowledge concept coverage feature $\boldsymbol{\tau}_{C_l}$.
The concatenated embedding is formulated as $\textbf{e}_{C_l}=[\textbf{x}_s, \textbf{x}_{C_l}, \boldsymbol{\tau}_{C_l} ] \in \mathbb{R} ^{q_s + q_e + m}$, where $q_s$ and $q_e$ denote the embedding size of student $s$ and exercise $C_l$, respectively, and $\boldsymbol{\tau}_{C_l}^k \in \{0, 1\}$ is binary meaning that exercise $C_l$ contains knowledge concept $k$ iff $\boldsymbol{\tau}_{C_l}^k = 1$.
The forward output state $\overrightarrow{\textbf{h}}_{C_l}=LSTM(\textbf{e}_{C_l},\textbf{h}_{C_{l-1}} )$ and the backward output state $\overleftarrow{\textbf{h}}_{C_l}=LSTM(\textbf{e}_{C_l},\textbf{h}_{C_{l+1}} )$ of the $l$-th exercise can be acquired by a Bi-LSTM cell.
The final exercise relevance representation $\mathbf{h}_{C_l}$ is then obtained by concatenating the forward and backward hidden states: $\textbf{h}_{C_l}=[ \overrightarrow{\textbf{h}}_{C_l},\overleftarrow{\textbf{h}}_{C_l} ] \in \mathbb{R} ^{2 \times q_h}$, where $q_h$ is the hidden size.
Finally, all exercise relevance representations $\mathbf{h}_{C_l}$ for $1 \leq l \leq L$ are stacked to form a matrix $\mathbf{H}_C \in \mathbb{R}^{2 \times L \times q_h}$.
Consequently, the exercise relevance estimator is capable of modeling the relationship between each candidate exercise and each knowledge concept in the candidate subset for the student.
As mentioned in \cite{LiuXQ2023}, Bi-LSTM models \cite{SchP1997} can be replaced by other deep learning models that supports bidirectional sequences, such as Transformer \cite{Vas2017}, to capture interactions across exercises.

\subsubsection{Learning Pattern Diversity}
\looseness=-1
The learning pattern diversity estimator further adjusts recommendation diversity in the candidate set $C$ according to distinct learning pace of each student.
Some students require recommendation lists that cover diverse knowledge concepts, while others prefer exercises focused on several specific knowledge concepts.
As exercises from various knowledge concepts are intermingled in students' exercise records, we divide the entire long exercise sequence into separate sequences for each knowledge concept to more effectively model the student's learning pace, as illustrated in Figure \ref{fig:work flow chart}.

\looseness=-1
To model intra-knowledge concept interactions, we employ LSTM to explicitly capture the time dependencies within each student's record for a given knowledge concept.
The final output state of the LSTM serves as a representation that encodes all information from the sequence and plays a pivotal role in subsequent predictions.
By separately encoding students’ learning patterns for each knowledge concept, we utilize self-attention mechanism \cite{Vas2017} to aggregate KC-specific signals and the learning pace distribution over knowledge concept.
We put all knowledge concept representation vectors into together $[\mathbf{k}_1, \cdots, \mathbf{k}_m]$ and obtain a matrix $\mathbf{W} \in \mathbb{R}^{m \times q_h}$ with $q_h$ as the hidden size.
Then we use multi-layer perceptron $f_m( \cdot )$ to generate the diversified learning pace distribution $ \boldsymbol{\hat{\omega}} \in \mathbb{R}^m$ over knowledge concepts.
    \begin{equation} \label{equation:self-attention}
        \begin{aligned}
             \boldsymbol{\hat{\omega}} = f_{m}(\text{Attention}(\mathbf{W})),
        \end{aligned}
    \end{equation}
where $\boldsymbol{\hat{\omega}}$ describes how likely a student is interested in each knowledge concept.
Moreover, the diversity of a given exercise within the candidate set $C$ is determined by its dissimilarity or novelty between the current exercise and the other exercises in $C$.
To quantify diversity, we adopt the probabilistic coverage function $b(\cdot)$, a widely used submodular function capturing diversity in recommendation tasks \cite{YueG2011}.
    \begin{equation} \label{equation:probability of at least one exercise}
        \begin{aligned}
             b_k(C) = 1 - \prod_{e \in C}{(1 - \tau_{e}^{k})},
        \end{aligned}
    \end{equation}
where $\tau_{e}^{k}$ is the knowledge concept coverage of exercise $e$, and $b_k(C)$ describes the probability of at least one exercise in $C$ covers knowledge concept $k$.

\looseness=-1
Additionally, the reduction in diversity is the decrease in knowledge concept coverage when removing exercise $C_l$ from $C$.
Hence, the marginal diversity of the $l$-th exercise in a candidate list $C$ is defined as:

    \begin{equation} \label{equation:diversity each KC}
        \begin{aligned}
             \mathbf{d}(C_l) = \textbf{b}(C) - \textbf{b}(C'),
        \end{aligned}
    \end{equation}
where $C' = C \setminus \{ C_l \}$, $\textbf{b}(C) = [b_1(C), \cdots, b_m(C)]$ and $\mathbf{d}(C_l) \in [0, 1]^m$.

\looseness=-1
Then, we obtain the learning pace diversity representation vector $\boldsymbol{\varDelta}(C_l)$ of each exercise $C_l$ where $1 \leq l \leq L$ from candidate list $C$.

    \begin{equation} \label{equation:deiversity representation}
        \begin{aligned}
             \boldsymbol{\varDelta}(C_l) =\boldsymbol{\hat{\omega}}\odot \mathbf{d}(C_l),
        \end{aligned}
    \end{equation}
where $\odot$ denotes the Hadamard product. 
The $k$-th element of $\boldsymbol{\varDelta} ( C_l)$ indicates the learning pace diversity gain of exercise $C_l$ for knowledge concept $k$.

\subsubsection{Re-ranking}
We employ two methods: deterministic and probabilistic to compute the final re-ranking scores.

\looseness=-1
\begin{enumerate}
	\item Deterministic Method: We stack the learning pace diversity representation vectors $\boldsymbol{\varDelta}(C_l)$ where $1 \leq l \leq L$ into a matrix $\boldsymbol{\varDelta}(C)\in \mathbb{R}^{L\times m}$.
	We then concatenate the exercise relevance matrix $\mathbf{H}_C$ and the learning pattern diversity matrix $\boldsymbol{\varDelta}(C)$ to get $\mathbf{V}$.
	Furthermore, we use an MLP($f_m(\cdot)$) to fuse the relevance and diversity in order to predict scores,
	\begin{equation} \label{equation:deterministic approach}
		\begin{aligned}
			\phi(C) = f_{m_\phi}(\mathbf{V}),
		\end{aligned}
	\end{equation}
	where $\phi(C)\in \mathbb{R} ^L$ and $\phi_l(C)$ is the $l$-th element of $\phi(C)$ representing the probability of exercise $C_l$ being suitable for the students.
	Finally, we sort the candidate list by $\phi_l(C)$, and select the top-k exercises as re-ranked list $R$ for the student.
	
	\item Probabilistic Method: We separately estimate a mean $f_{m_\varphi}( \mathbf{V} )$ and a standard deviation $f_{m_s}( \mathbf{V} )$ for the re-ranking score of each exercise \cite{LiuXQ2023},
	
	\begin{equation} \label{equation:incorporating variable}
		\begin{aligned}
			\phi_l(C) = f_{m_\varphi}(\mathbf{V}) + \xi_{C_l}{f_{m_s}( \mathbf{V})},
		\end{aligned}
	\end{equation}
	where $f_{m_\varphi}(\mathbf{V}) \in \mathbb{R} ^L$, $f_{m_s}(\mathbf{V}) \in \mathbb{R}^L$ and $\xi_{C_l} \sim \mathscr{N}(0, 1)$.
	The $\phi_l(C)$ denotes the re-ranking score for the exercise $C_l$ with $1 \leq l \leq L$.
	In training, we transform the stochastic sampling into Equation \eqref{equation:incorporating variable} by incorporating standardized normal random variables.
	At the inference phase, we use the confidence bound $U(C_l)$ of the re-ranking score for exercise $C_l$ to perform re-ranking task.
	
	\begin{equation} \label{equation:re-ranking scores}
		\begin{aligned}
			U(C_l) = f_{m_\varphi}(\mathbf{V}) + f_{m_s}(\mathbf{V}).
		\end{aligned}
	\end{equation}
	
	\looseness=-1
	The standard deviation $f_{m_s}(V)$ describes the confidence of the model in the current estimated re-ranking scores and aims to promote exploration, which is also tailored to individual students.
	Similarly, we derive the re-ranked list by choosing the top-k exercises based on the upper confidence bound of the re-ranking scores $U( C_l)$ to maximize the matching learning pace of the student.
\end{enumerate}

\subsubsection{Optimization}
\looseness=-1
We minimize the cross-entropy loss so as to ensure that the model assigns a higher score to exercises that the student has not mastered compared to those that have been mastered.
    \begin{equation} \label{optimization}
        \begin{aligned}
             \mathcal{L} =-\sum_{l=1}^L{\{ y_llog( \phi( C_l  ) ) +  ( 1-y_l) log( 1-\phi( C_l  ) ) \}},
        \end{aligned}
    \end{equation}
where $y_l \in \{ 0,1\}$ is the indicator for the $l$-th exercise in the subset $C$, with $1$ denoting an exercise that is not mastered and $0$ denoting an exercise that is mastered.
Following this approach, the neural re-ranking module is able to learn the tradeoff between exercise relevance and learning pattern diversity in an end-to-end manner, directly from student feedback.

\section{Experiments}
\looseness=-1
We evaluate the performance of NR4DER in answering the following research questions (RQs):
\begin{description}
\item[RQ1:] How does NR4DER perform in various recommendation scenarios compared to state-of-the-art exercise recommendation models?
\item[RQ2:] Does the student representation enhancer module effectively alleviate the long-tail problem for active and inactive students?
\item[RQ3:] How does neural re-ranking provide diverse exercises based on the varying learning pace of students?
\end{description}

    \begin{table}[!t]
     \caption{Dataset description of three datasets}
    \centering
    \begin{tabular}{lcccc}
    \toprule
    Dataset        & Students & KCs  & Exercises & Interactions \\ \hline
    Nips34         & 4,918     & 57   & 984       & 1,399,470    \\ 
    Assist2009 & 4,217     & 123  & 17,737     & 374,422    \\ 
    Assist2012         & 27,066     & 265   & 45,716       & 2,541,201 \\
    \bottomrule
    \end{tabular}
    \label{table:dataset_info}
    \end{table}
    
\subsection{Datasets}
\looseness=-1
We conducted experiments with three  widely used educational datasets to validate the effectiveness of our model.
The details about the datasets are shown in Table \ref{table:dataset_info}.

\begin{itemize}
\item Nips34 is provided by the NeurIPS 2020 Education Challenge and contains answers to math questions from Eedi \cite{PieBH2015}. We used the datasets from Tasks 3 and 4 to evaluate our model.
\item Assist2009 focuses on math exercises, with data from ASSISTments \cite{FenHK2009}, a free online tutoring platform, for the period 2009-2010. 
\item Assist2012 is from the ASSISTments platform, spanning the 2012-2013 period, and includes affect predictions.
\end{itemize}

 \begin{table*}
    \caption{Comparative experiments across  three datasets. Improvements of NR4DER vs. baseline models are measured in Overall and Mean performance based on average of metrics. Bold indicates the best performance among all methods, while underline indicates the second-best performance.}
    \centering
    \footnotesize
    \resizebox{\textwidth}{!}{ 
        \begin{tabular}{c|c|*{12}{>{\centering\arraybackslash}p{0.9cm}}}
        \hline
        Dataset & Model & NDCG@1 & NDCG@3 & NDCG@5 & NDCG@10 & F1@1 & F1@3 & F1@5 & F1@10 & Recall@1 & Recall@3 & Recall@5 & Recall@10 \\
        \hline
        \multirow{8}{*}{Nips34} & DKTRec & 0.697 & 0.822 & 0.830 & 0.824 & 0.697 & 0.736 & 0.726 & 0.688 & 0.697 & 0.650 & 0.620 & 0.567 \\
        & DKVMNRec & 0.689 & 0.826 & 0.830 & 0.825 & 0.689 & 0.738 & 0.727 & 0.689 & 0.689 & 0.652 & 0.623 & 0.567 \\
        & AKTRec & 0.740 & 0.853 & 0.855 & 0.846 & 0.740 & 0.768 & 0.749 & 0.700 & 0.740 & 0.688 & 0.648 & 0.582 \\
        & SimpleKTRec & 0.735 & 0.852 & 0.854 & 0.845 & 0.735 & 0.772 & 0.749 & 0.702 & 0.735 & 0.692 & 0.650 & 0.584 \\
        & KCPER & 0.338 & 0.545 & 0.609 & 0.646 & 0.338 & 0.488 & 0.541 & 0.590 & 0.338 & 0.401 & 0.432 & 0.465 \\
        & KG4Ex & 0.486 & 0.693 & 0.731 & 0.732 & 0.486 & 0.590 & 0.608 & 0.622 & 0.486 & 0.486 & 0.487 & 0.495 \\
        & MMER & 0.472 & 0.633 & 0.666 & 0.686 & 0.472 & 0.631 & 0.627 & 0.625 & 0.472 & 0.461 & 0.456 & 0.454 \\
        \rowcolor{gray!20}
        & \textbf{NR4DER-p} & \textbf{0.915} & \textbf{0.939} & \textbf{0.934} & \textbf{0.919} & \textbf{0.915} & \textbf{0.879} & \textbf{0.828} & \underline{0.727} & \textbf{0.915} & \textbf{0.839} & \textbf{0.762} & \underline{0.622} \\
        \rowcolor{gray!20}
        & \textbf{NR4DER-d} & \underline{0.905} & \underline{0.929} & \underline{0.924} & \underline{0.914} & \underline{0.905} & \underline{0.863} & \underline{0.822} &  \textbf{0.729} & \underline{0.905} & \underline{0.821} & \underline{0.756} & \textbf{0.628}  \\
        \hline
        \multirow{8}{*}{Assist2009} & DKTRec & 0.543 & 0.674 & 0.709 & 0.726 & 0.543 & 0.475 & 0.434 & 0.385 & 0.543 & 0.419 & 0.367 & 0.311 \\
        & DKVMNRec & 0.546 & 0.673 & 0.701 & 0.716 & 0.546 & 0.470 & 0.430 & 0.382 & 0.546 & 0.412 & 0.364 & 0.309 \\
        & AKTRec & \underline{0.552} & 0.581 & 0.719 & 0.732 & \underline{0.552} & \underline{0.480} & 0.437 & \underline{0.386} & \underline{0.552} & \underline{0.424} & \underline{0.369} & \underline{0.312} \\
        & SimpleKTRec & 0.543 & 0.667 & 0.701 & 0.713 & 0.543 & 0.469 & 0.435 & 0.385 & 0.543 & 0.411 & 0.367 & \underline{0.312} \\
        & KCPER & 0.216 & 0.291 & 0.311 & 0.297 & 0.216 & 0.282 & 0.304 & 0.320 & 0.216 & 0.241 & 0.251 & 0.261 \\
        & KG4Ex & 0.251 & 0.349 & 0.374 & 0.324 & 0.251 & 0.295 & 0.300 & 0.304 & 0.251 & 0.250 & 0.249 & 0.250 \\
        & MMER & 0.327 & 0.459 & 0.502 & 0.529 & 0.327 & \textbf{0.500} & \textbf{0.499} & \textbf{0.492} & 0.327 & 0.332 & 0.333 & \textbf{0.326} \\
        \rowcolor{gray!20}
        & \textbf{NR4DER-p} & \textbf{0.569} & \textbf{0.711} & \textbf{0.751} & \textbf{0.775} & \textbf{0.569} & \underline{0.480} & \underline{0.473} & 0.371 & \textbf{0.569} & \textbf{0.425} & \textbf{0.369} & 0.302 \\
        \rowcolor{gray!20}
        & \textbf{NR4DER-d} & 0.546 & \underline{0.702} & \underline{0.740} & \underline{0.761} & 0.546 & 0.471 & 0.425 &  0.370 & 0.546 & 0.414 & 0.359 & 0.301  \\
        \hline
        \multirow{9}{*}{Assist2012} 
        & DKT          & 0.347   & 0.499   & 0.528   & 0.537   & 0.347   & 0.385   & 0.345   & 0.26    & 0.347   & 0.298   & 0.247   & 0.171   \\
        & DKVMN        & 0.353   & 0.497   & 0.528   & 0.537   & 0.353   & 0.38    & 0.342   & 0.259   & 0.353   & 0.294   & 0.244   & 0.17    \\
        & AKT          & 0.368   & 0.516   & 0.542   & 0.549   & 0.368   & 0.396   & 0.353   & 0.263   & 0.368   & 0.308   & 0.254   & 0.173   \\
        & SimpleKT     & 0.385   & 0.52    & 0.547   & 0.553   & 0.385   & 0.396   & 0.354   & 0.263   & 0.385   & 0.309   & 0.254   & 0.173   \\
        & KCPER        & 0.143   & 0.287   & 0.346   & 0.411   & 0.143   & 0.265   & 0.308   &  0.332   & 0.143   & 0.212   & 0.239   & 0.254   \\
        & KG4Ex        & 0.355   & 0.375   & 0.489   & 0.592   & 0.355   & 0.368   &  0.379 & 0.415 & 0.355   & 0.323   & 0.265 & 0.197   \\
        & MMER         & 0.315   & 0.506   & 0.562 & 0.603 & 0.315   & 0.399 & 0.417 & 0.434 & 0.315   & 0.314 & 0.312 & 0.307 \\
        \rowcolor{gray!20}
        & \textbf{NR4DER-p}   & \textbf{0.510} & \textbf{0.655} & \textbf{0.734} & \textbf{0.816} & \textbf{0.510} & \textbf{0.506} & \underline{0.500} & \textbf{0.470} & \textbf{0.510} & \textbf{0.418} & \underline{0.386} & \textbf{0.336} \\
        \rowcolor{gray!20}
        & \textbf{NR4DER-d}  & \underline{0.491} & \underline{0.649} & \underline{0.721} & \underline{0.806} & \underline{0.491} & \underline{0.505} & \textbf{0.503} & \underline{0.463} & \underline{0.491} & \underline{0.415} & \textbf{0.389} & \underline{0.330} \\

        \hline
        \end{tabular}
        }
        \label{overall performance}
       \end{table*}

\begin{table*}
\caption{Ablation Experiment: "w/o En" represents the model without the sequence enhancement module, and "w En" represents the overall model.}
\centering
\footnotesize
\resizebox{\textwidth}{!}{ 
\begin{tabular}{c|c|*{12}{>{\centering\arraybackslash}p{0.9cm}}}

\hline\hline
Dataset                     & Model             & NDCG@1         & NDCG@3         & NDCG@5         & NDCG@10        & F1@1           & F1@3           & F1@5           & F1@10          & Recall@1       & Recall@3       & Recall@5       & Recall@10      \\ \hline\hline
\multirow{6}{*}{Nips34}     & w/o En(overall)   & 0.854          & 0.901          & 0.898          & 0.882          & 0.854          & 0.820          & 0.782          & 0.707          & 0.854          & 0.765          & 0.702          & 0.596          \\
                            & w En(overall)     & \textbf{0.915} & \textbf{0.939} & \textbf{0.934} & \textbf{0.919} & \textbf{0.915} & \textbf{0.879} & \textbf{0.828} & \textbf{0.727} & \textbf{0.915} & \textbf{0.839} & \textbf{0.762} & \textbf{0.622} \\ \cline{2-14} 
                            & w/o En(active stu.) & 0.910          & 0.925          & 0.920          & 0.906          & 0.910          & 0.865          & 0.824          & 0.727          & 0.910          & 0.827          & 0.763          & 0.630          \\
                            & w En(active stu.)   & \textbf{0.940} & \textbf{0.946} & \textbf{0.942} & \textbf{0.935} & \textbf{0.940} & \textbf{0.911} & \textbf{0.871} & \textbf{0.757} & \textbf{0.940} & \textbf{0.889} & \textbf{0.831} & \textbf{0.675} \\ \cline{2-14} 
                            & w/o En(inactive stu.) & 0.843          & 0.883          & 0.885          & 0.879          & 0.843          & 0.786          & 0.757          & 0.705          & 0.843          & 0.715          & 0.663          & 0.587          \\
                            & w En(inactive stu.)   & \textbf{0.880} & \textbf{0.918} & \textbf{0.917} & \textbf{0.901} & \textbf{0.880} & \textbf{0.821} & \textbf{0.779} & \textbf{0.716} & \textbf{0.880} & \textbf{0.759} & \textbf{0.690} & \textbf{0.601} \\ \hline\hline
\multirow{6}{*}{Assist2009} & w/o En(overall)   & 0.543          & 0.670          & 0.706          & 0.720          & 0.513          & 0.461          & 0.421          & 0.367          & 0.513          & 0.404          & 0.354          & 0.291          \\
                            & w En(overall)     & \textbf{0.569} & \textbf{0.711} & \textbf{0.751} & \textbf{0.775} & \textbf{0.569} & \textbf{0.480} & \textbf{0.430} & \textbf{0.371} & \textbf{0.569} & \textbf{0.425} & \textbf{0.369} & \textbf{0.302} \\ \cline{2-14} 
                            & w/o En(active stu.) & 0.550          & 0.731          & 0.761          & 0.757          & 0.550          & 0.478          & 0.434          & 0.347          & 0.550          & 0.410          & 0.353          & 0.261          \\
                            & w En(active stu.)   & \textbf{0.584} & \textbf{0.773} & \textbf{0.801} & \textbf{0.787} & \textbf{0.584} & \textbf{0.502} & \textbf{0.445} & \textbf{0.351} & \textbf{0.584} & \textbf{0.438} & \textbf{0.367} & \textbf{0.268} \\ \cline{2-14} 
                            & w/o En(inactive stu.) & 0.515          & 0.644          & 0.693          & 0.706          & 0.515          & 0.454          & 0.416          & 0.382          & 0.515          & 0.390          & 0.350          & 0.316          \\
                            & w En(inactive stu.)   & \textbf{0.560} & \textbf{0.686} & \textbf{0.713} & \textbf{0.740} & \textbf{0.560} & \textbf{0.475} & \textbf{0.429} & \textbf{0.385} & \textbf{0.560} & \textbf{0.422} & \textbf{0.367} & \textbf{0.320} \\ \hline\hline
\multirow{6}{*}{Assist2012} & w/o En(overall) & 0.502 & 0.649 & 0.724 & 0.812 & 0.502 & 0.458 & 0.419 & 0.379 & 0.502 & 0.381 & 0.331 & 0.287 \\ 
                            & w En(overall) & \textbf{0.510} & \textbf{0.655} & \textbf{0.734} & \textbf{0.816} & \textbf{0.510} & \textbf{0.506} & \textbf{0.500} & \textbf{0.470} & \textbf{0.510} & \textbf{0.418} & \textbf{0.386} & \textbf{0.336} \\ \cline{2-14} 
                            & w/o En(active stu.) & 0.512 & 0.653 & 0.737 & 0.813 & 0.512 & 0.468 & 0.464 & 0.442 & 0.512 & 0.408 & 0.344 & 0.308 \\ 
                            & w En(active stu.) & \textbf{0.522} & \textbf{0.667} & \textbf{0.745} & \textbf{0.829} & \textbf{0.522} & \textbf{0.512} & \textbf{0.511} & \textbf{0.480} & \textbf{0.522} & \textbf{0.435}  & \textbf{0.398} & \textbf{0.325} \\ \cline{2-14} 
                            & w/o En(inactive stu.) & 0.494 & 0.647 & 0.722 & 0.808 & 0.494 & 0.450 & 0.407 & 0.371 & 0.494 & 0.371 & 0.319 & 0.281 \\ 
                            & w En(inactive stu.) & \textbf{0.509} & \textbf{0.652} & \textbf{0.731} & \textbf{0.811} & \textbf{0.509} & \textbf{0.451} & \textbf{0.411} & \textbf{0.372} & \textbf{0.509} & \textbf{0.374}  & \textbf{0.322} & \textbf{0.282} \\ \hline\hline
\end{tabular}
}
\label{ablation experiment}
\end{table*}

\begin{table*}[ht]
\caption{Ablation Experiment: "w/o NR" represents the model without the neural re-ranking module, and "w NR" represents the overall model.}
\centering
\footnotesize
\resizebox{\textwidth}{!}{ 
\begin{tabular}{c|c|*{12}{>{\centering\arraybackslash}p{0.9cm}}}

\hline
Dataset                     & Model             & NDCG@1         & NDCG@3         & NDCG@5         & NDCG@10        & F1@1           & F1@3           & F1@5           & F1@10          & Recall@1       & Recall@3       & Recall@5       & Recall@10      \\ \hline
\multirow{2}{*}{Nips34}     & w/o NR & 0.511          & 0.598          & 0.608          & 0.613          & 0.511          & 0.692          & 0.718          & 0.726          & 0.511          & 0.501          & 0.495          & 0.487          \\
                           & w NR   & \textbf{0.915} & \textbf{0.939} & \textbf{0.934} & \textbf{0.919} & \textbf{0.915} & \textbf{0.879} & \textbf{0.828} & \textbf{0.727} & \textbf{0.915} & \textbf{0.839} & \textbf{0.762} & \textbf{0.622} \\ \hline  
\multirow{2}{*}{Assist2009} & w/o NR & 0.271          & 0.313          & 0.321          & 0.326          & 0.271          & 0.368          & 0.380          & \textbf{0.377} & 0.271          & 0.265          & 0.265          & 0.265          \\
                            & w NR   & \textbf{0.569} & \textbf{0.711} & \textbf{0.751} & \textbf{0.775} & \textbf{0.569} & \textbf{0.480} & \textbf{0.430} & 0.371          & \textbf{0.569} & \textbf{0.425} & \textbf{0.369} & \textbf{0.302} \\  \hline 
\multirow{2}{*}{Assist2012} 
& w/o NR    & 0.281 & 0.346 & 0.354 & 0.362 & 0.281 & 0.454 & 0.491 & \textbf{0.531} & 0.281 & 0.273 & 0.269 & 0.272  \\
& w NR       & \textbf{0.510} & \textbf{0.655} & \textbf{0.734} & \textbf{0.816} & \textbf{0.510} & \textbf{0.506} & \textbf{0.500} & \underline{0.470} & \textbf{0.510} & \textbf{0.418} & \textbf{0.386} & \textbf{0.336}  \\ \hline      
\end{tabular}
}
\label{ablation experiment on re-ranking}
\end{table*}

\subsection{Baselines}
In order to verify the effectiveness of our model, our model is compared with 7 exercise recommendation algorithms respectively:  
	
\begin{itemize}
	\item DKTRec \cite{PieBH2015} first introduces deep learning into KT and leverage LSTM to efficiently model learning process.
	\item DKVMNRec \cite{ZhaSK2017} leverages a key-value memory network that is able to utilize the relationships between underlying concepts to directly output students' mastery of each knowledge concept.		
	\item AKTRec \cite{GhoHL2020} is an attention-based knowledge tracing approach that summarizes learners' past performance.
	\item SimpleKTRec \cite{LiuLC2023} captures individual differences among students and uses dot product attention to simplify the modeling of learning interactions.
	\item KCPER \cite{WuLT2020} combines LSTM and DKT to predict students' knowledge states and uses simulated annealing algorithm to enhance recommendation diversity.
    \item KG4Ex \cite{GuaXC2023} applies a knowledge graph to enhance exercise recommendation interpretability.
    \item MMER \cite{LiuHL2023} views knowledge concepts as multi-agent that both compete and cooperate, enhancing recommendation robustness for new students through meta-training.
\end{itemize}

\subsection{Evaluation Metrics}
\looseness=-1
In order to comprehensively assess the recommendation performance, we use three commonly used metrics in our experiments: normalised discounted cumulative gain (NDCG), recall (Recall), F1-score (F1), diversity (DIV) \cite{LiuXQ2023}.
The higher values of these metrics represent the better recommendation performance. In addition, we calculate Metrics@K (K = 1, 3, 5, 10) based on the length of the exercise lists to test the effectiveness of the recommendation results under different lengths of recommendation lists.
This multi-dimensional evaluation approach ensures that we have a comprehensive understanding of the performance of NR4DER.

\subsection{Implementation Details}
\looseness=-1
Our method is implemented in PyTorch on NVIDIA RTX 4090 with 24 GB of memory.
Based on the recommended values, we refine the initial hyperparameters and proceed with experiments utilizing the specified parameters listed below: $\beta$ is set to $\{0.4, 0.6, 0.8, 1.0\}$, $\gamma_s$ is set to $\{0.1, 0.3, 0.5, 0.7, 1.0\}$, $\delta$ is 0.7 \cite{WuLT2020}, the learning rate is $0.001$, the batch size is $\{16, 32, 64, 128\}$, and the number of heads of multi-head attention is $\{2, 4, 6, 8\}$, the ratio of the training set to the test set is $8:2$.
Our code is available at \url{https://github.com/chanllon/NR4DER.Exercise-Recommendation}.

\subsection{(RQ1) Performance Comparison}
\looseness=-1
To validate the effectiveness of NR4DER, we conduct a comparative analysis on the three datasets mentioned above.
Table \ref{overall performance} presents the experimental results of NR4DER in comparison to baseline models.
In most metrics, NR4DER outperforms the other baseline models. 
However, NR4DER exhibits relatively lower performance on the F1 metric for the Assist2009.
This is because the F1 metric focuses on classification accuracy rather than ranking quality, while NR4DER emphasizes ranking quality of recommendation results.
While NR4DER exhibits marginal improvement in the F1 metric, it maintains superior overall performance across comprehensive evaluations.
The result suggests that NR4DER, by introducing the student representation enhancer, effectively improves the representation of learning sequences for inactive students, enabling the model to better capture their learning pace.
Additionally, through the neural re-ranking module, the model can extract and capture the deep interrelations between knowledge concepts from the historical learning records of students, thereby recommending more suitable exercises that encompass diverse knowledge concepts and significantly improving the overall performance of exercise recommendation.

\begin{figure*}[ht]
    \centering
    \subfigure[Diversity of Nips34]{
        \includegraphics[width=0.32\linewidth]{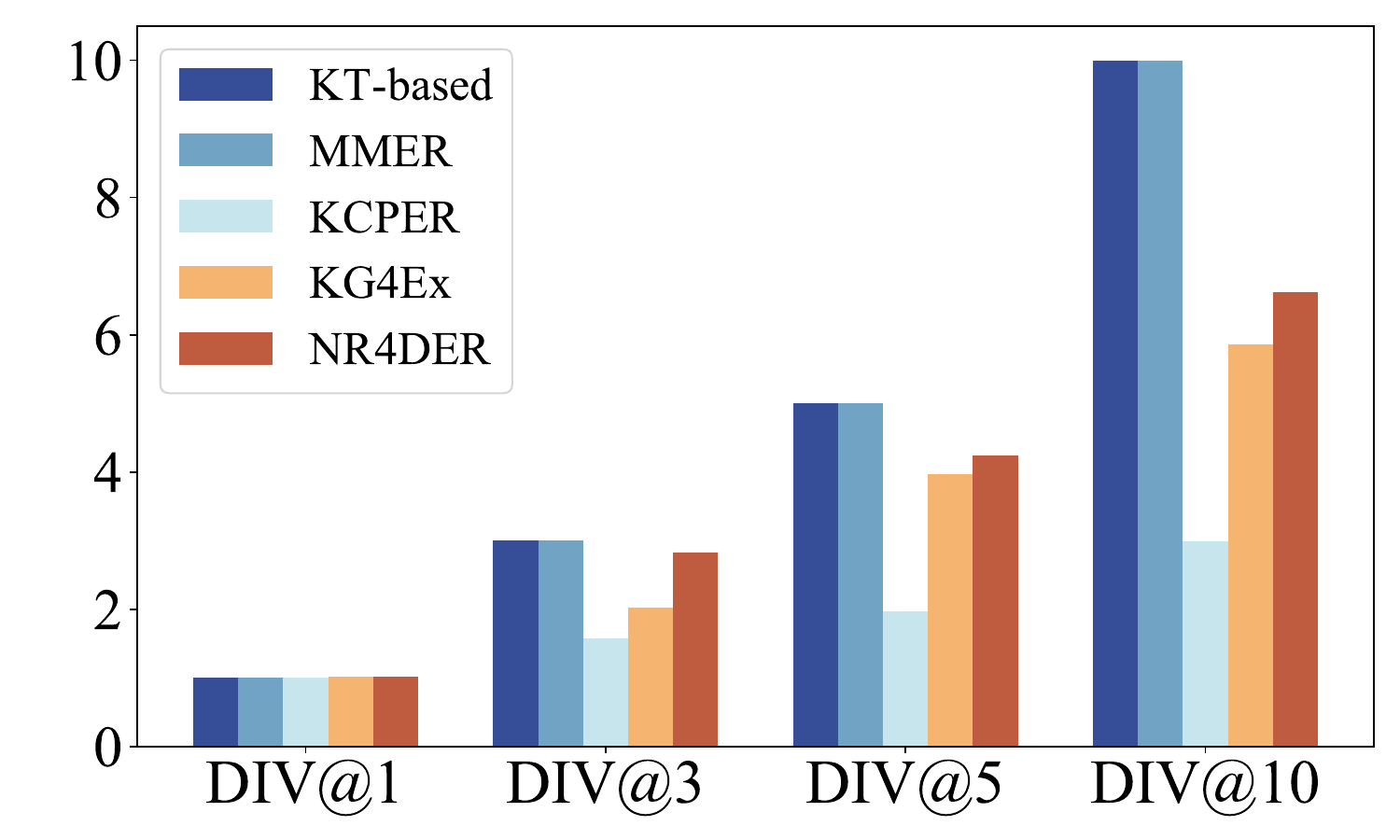}
    }
    \hfill
    \subfigure[Diversity of Assist2009]{
        \includegraphics[width=0.32\linewidth]{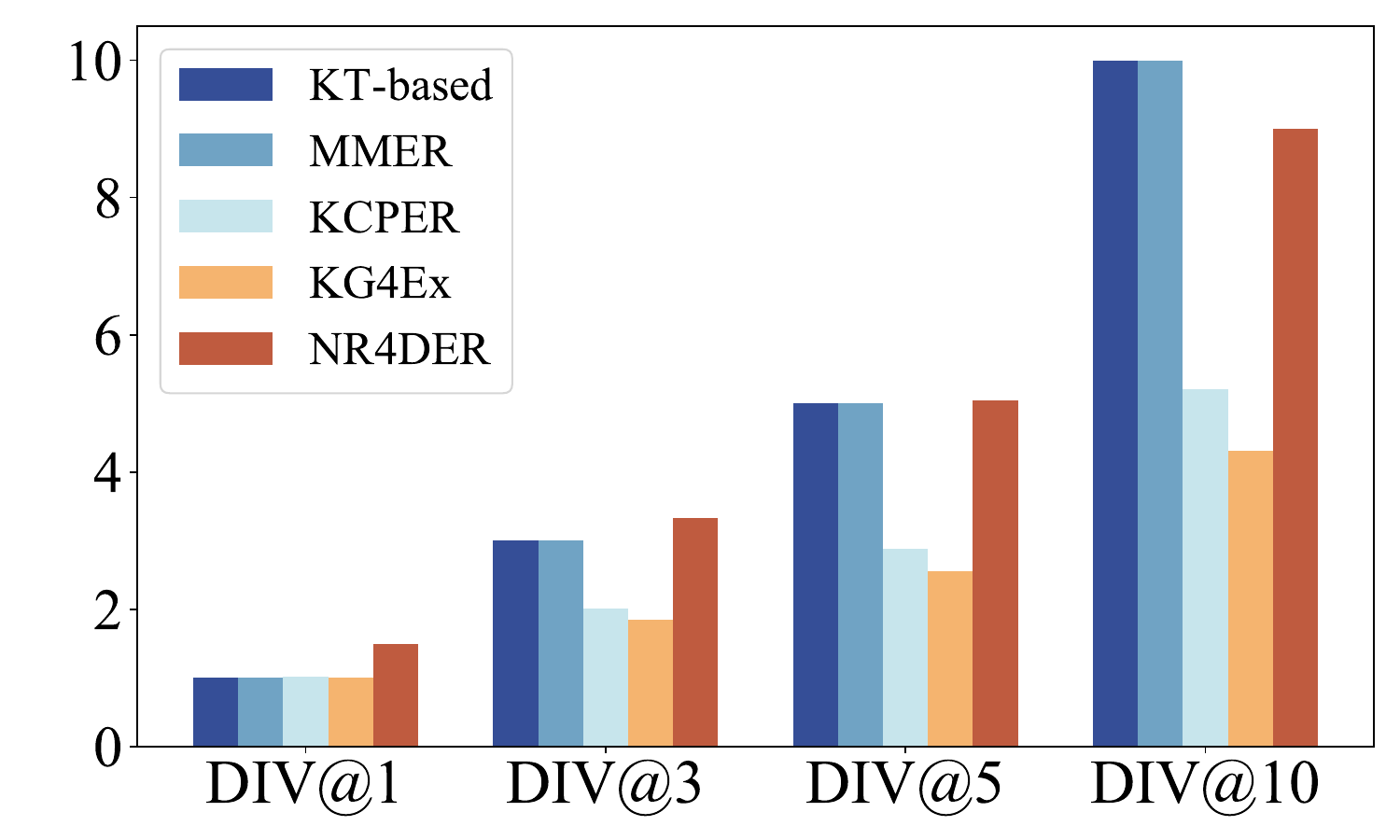}
    }
    \hfill
    \subfigure[Diversity of Assist2012]{
        \includegraphics[width=0.32\linewidth]{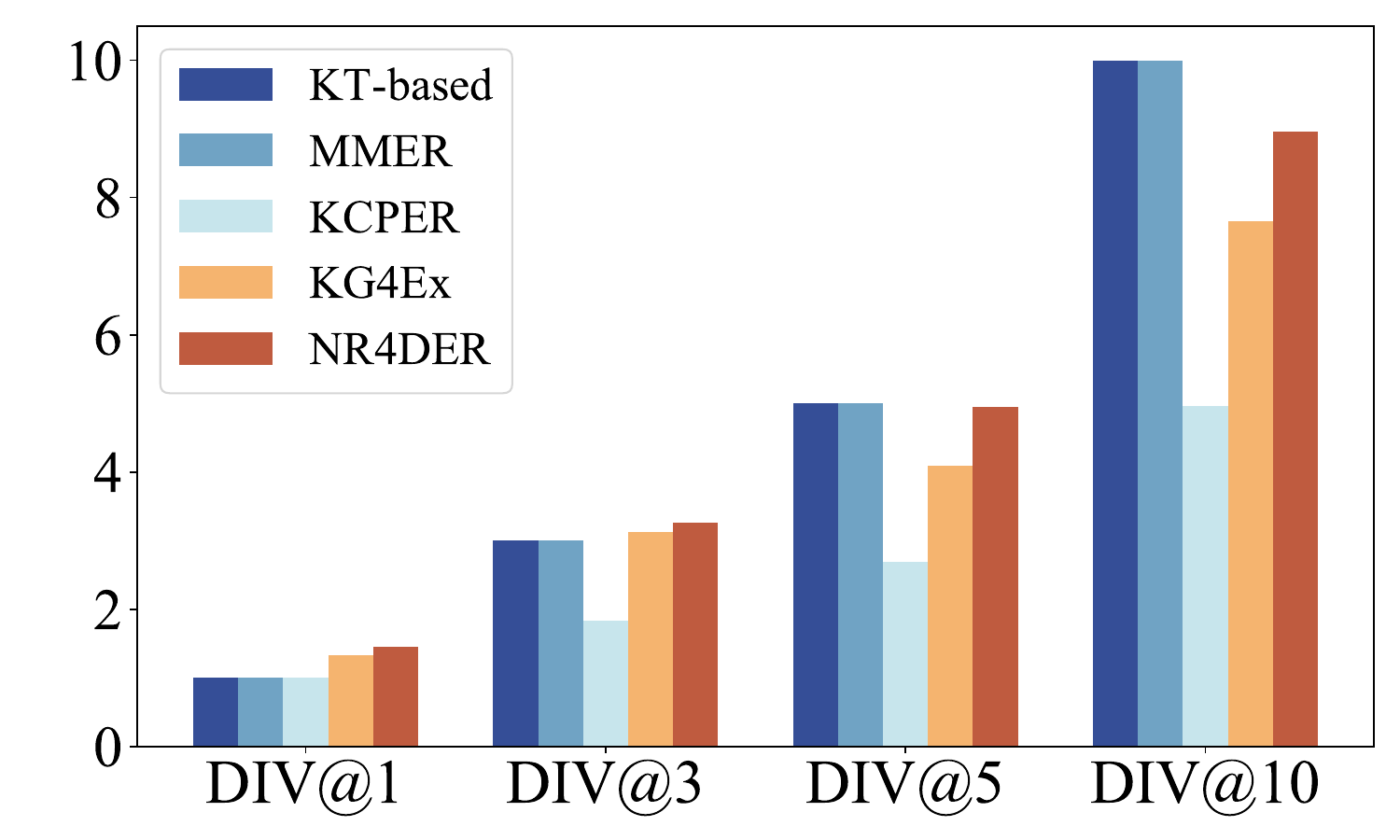}
    }
    \caption{The diversity of three datasets.}
    \Description{The diversity of three datasets.}
    \label{fig:div}
\end{figure*}

\begin{figure}[ht]
	\centering
	\includegraphics[width= 1\linewidth]{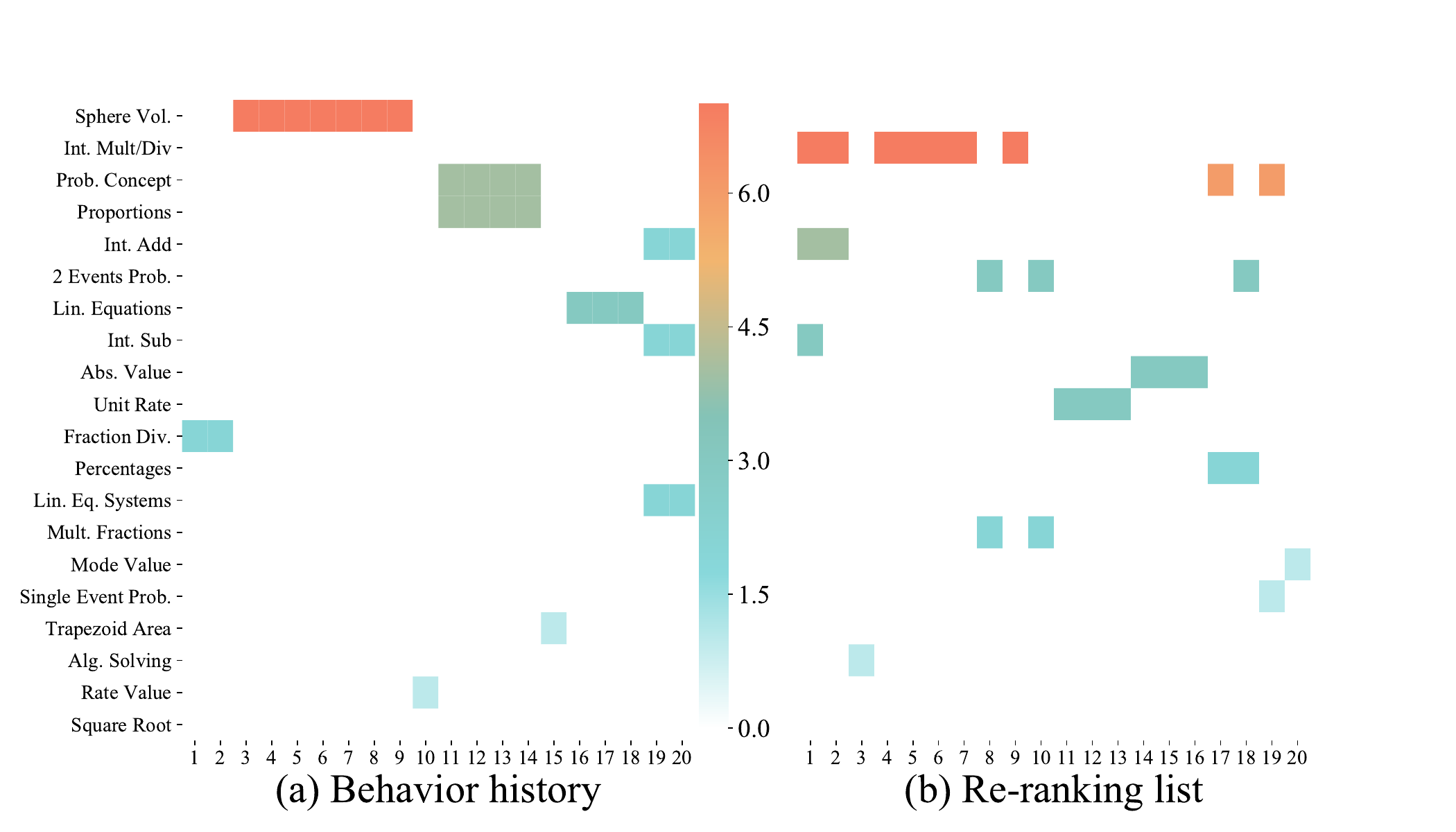}
	\caption{The distribution of knowledge concepts.}
	\label{fig:case}
\end{figure}

\subsection{(RQ2) Ablation Experiment}
\looseness=-1
We conduct ablation experiments to evaluate the impact of the student representation enhancer and the neural re-ranking module on the overall effectiveness of the system.

\looseness=-1
As for the student representation enhancer module, students were divided into active (top 5\% of interactions) and inactive (the remaining 95\%) groups for separate performance comparison. 
As shown in Table \ref{ablation experiment}, all three datasets clearly exhibit a long-tail problem.
The results demonstrate that the full model significantly outperforms the model without the student representation enhancer across multiple metrics, highlighting its crucial role in improving overall model performance.
Specifically, the student representation enhancer substantially improves ability of the model so as to capture the learning behavior of inactive students.
When this module is removed, the model struggles to adequately capture the learning patterns, resulting in a decline in recommendation accuracy.

\looseness=-1
Figure 5 shows the effectiveness of neural re-ranking module in NR4DER.
The experimental results demonstrate that neural re-ranking module brings performance improvement over the backbone model across all datasets. 
In addition, we find that the neural re-ranking module shows a significant performance improvement on the NDCG and Recall metrics.
This indicates that the neural re-ranking module prioritizes the ranking quality of the recommendation list, leading to better-ranked exercises that align more closely with the learning pace of students.
Overall, NR4DER demonstrates excellent performance and substantiate the effectiveness of the neural re-ranking module.

\subsection{(RQ3) Diversity Analysis}
\looseness=-1
To comprehensively evaluate the diversity of NR4DER in generating exercise recommendations, we introduce a diversity metric.
Notably, KT-based methods and MMER models generate static recommendation lists of exercises with non-mastered knowledge concepts \cite{LiuHL2023}.
In addition, each exercise involves distinct knowledge concept.
Hence, these methods exhibit high and stable diversity.
However, these methods fail to adequately account for the diverse learning pace of students, resulting in lower accuracy compared to NR4DER, as shown in Table \ref{overall performance}.
Furthermore, as shown in Figure \ref{fig:div}, NR4DER consistently outperforms KCPER and KG4Ex across diversity evaluation metrics.
Although KCPER aims to maximize the diversity of exercises related to different knowledge concepts in the recommendation list, it often leads to similar re-ranked results, hence overlooking the learning pace of students and limiting diversity.
Furthermore, KG4Ex focuses primarily on the explainability of exercise recommendations and lacks the diverse learning paces of students, hence negatively impacting both performance and diversity.
To sum up, these demonstrate that our model not only provides suitable exercises but also significantly enhances the diversity of the recommended content.

\looseness=-1
As shown in Figure \ref{fig:case}, to illustrate how NR4DER matches the learning pace of students, we randomly select a student from Assist2009 and visualize the distribution of knowledge concepts in historical exercises alongside the 20 exercises
The brighter the color in the figure, the more frequently the knowledge concept has been studied.
NR4DER effectively explores combinations of knowledge concepts from the student’s historical exercises and recommends exercises that partially overlap with previously practiced knowledge concepts to reinforce prior learning.
For instance, after students complete integer addition/subtraction exercises, NR4DER prioritizes recommending problems integrating these mastered knowledge concepts with multiplication/division, hence reinforcing prior knowledge concepts.
Subsequently, NR4DER provide exercises involving diverse knowledge concepts.
Based on the knowledge state of the student, NR4DER may revisit previously attempted but not fully understood exercises, such as those involving probability concepts.
In summary, NR4DER effectively identifies combinations of related knowledge concepts and achieves diversified re-ranking based on the learning pace of students.

\section{Conclusion}
\looseness=-1
To match the diverse learning pace of students and address the long-tailed student distribution problem, we introduce the NR4DER model, a neural re-ranking method for diversified exercise recommendation.
NR4DER enhances the representation of student learning behaviors through a sequence enhancement module and optimizes the diversity and relevance of recommended exercises using neural re-ranking techniques.
Experimental results demonstrate that NR4DER performs exceptionally well on multiple real-world datasets, significantly improving the accuracy of recommendations while catering to students' diverse learning paces.

\begin{acks}
This paper was supported by National Key R\&D Program of China (2022YFC3303603), NSFC (62377028, 62477016, 62276114), Guangdong Basic and Applied Basic Research Foundation (2023B151512 0064), Key Laboratory of Smart Education of Guangdong Higher Education Institutes, Jinan University (2022LSYS003), Guangzhou Science and Technology Planning Project (Nansha District: 2023ZD001, Development District: 2023GH01, 2025A03J3565) and FDCT/0126/ 2024/RIA2.
\end{acks}

\bibliographystyle{ACM-Reference-Format}
\balance
\bibliography{main}

\end{document}